\documentclass[lettersize,journal]{IEEEtran}
\IEEEoverridecommandlockouts
\usepackage{tabularx}
\usepackage[T1]{fontenc}
\usepackage{algorithm}
\usepackage[noend]{algpseudocode}
\usepackage{graphicx}
\usepackage{amsmath,amssymb,amsfonts,stfloats}
\usepackage{array}
\usepackage{textcomp}

\usepackage{color}

\usepackage{paralist}
\usepackage{lettrine}
\usepackage{cite}
\usepackage{xcolor}
\usepackage{cancel}
\usepackage{mathtools}
\usepackage{bbm}
\usepackage{array}
\usepackage{booktabs}
\usepackage[caption=false,font=footnotesize]{subfig}
\usepackage{url}
\usepackage{acronym}
\usepackage{multicol}
\usepackage{makecell}
\usepackage{enumitem}
\usepackage{hyperref}

\hypersetup{colorlinks=true, citecolor=blue, linkcolor=blue, urlcolor=cyan}

\begin{document}

\title{Compressed LLM Reprogramming for Vision-Aided Beam Prediction in Vehicular Networks} 


\author{Kai Dong,~\IEEEmembership{Member,~IEEE,} 
Lei Wang,~\IEEEmembership{Graduate Student Member,~IEEE,} Changyi Li,~\IEEEmembership{Graduate Student Member,~IEEE,} 
Sergiy A. Vorobyov,~\IEEEmembership{Fellow,~IEEE} and Stefan Werner,~\IEEEmembership{Fellow,~IEEE} 
\thanks{This work is partially supported by the Research Council of Finland (Grant 354523).}

\thanks{The authors are with the Department of Information and Communications Engineering, Aalto University, 02150 Espoo, Finland (e-mail: kai.dong@aalto.fi,  lei.wang@aalto.fi, changyi.li@aalto.fi, sergiy.vorobyov@aalto.fi, and stefan.werner@aalto.fi)}

}
\maketitle

\begin{abstract}
Large Language Model (LLM) reprogramming-based beam prediction demonstrates strong data efficiency by adapting pretrained language models for vehicle-to-infrastructure (V2I) beam prediction, yet the resulting model complexity makes such approaches impractical for latency-sensitive deployment. 
We propose LLMBP-Lite, a compact LLM-reprogrammed framework for beam prediction. It leverages structural redundancy through pruning along three complementary dimensions: Transformer depth, source-prototype vocabulary size, and prompt length. Additionally, knowledge distillation can be optionally employed to maintain the pretrained representational capacity after compression.
Experiments on the real-world dataset demonstrate that LLMBP-Lite achieves a $65\times$ inference speedup over the uncompressed LLM-based baseline while maintaining prediction accuracy and consistently outperforming recurrent baselines under limited training data. These results demonstrate that the tradeoff between data efficiency and deployment efficiency can be substantially mitigated in vehicular networks.
\end{abstract}


\section{Introduction}
\label{sec:intro} 
\IEEEPARstart{T}{he} next-generation vehicular networks operating in high-frequency bands requires highly directional transmission, making beam management a challenging problem~\cite{dong2025network}.  In highly dynamic vehicular environments, accurate beam prediction is required to maintain reliable communication links, which may incur prohibitive overhead and high latency, especially when exhaustive beam search is adopted~\cite{xue2024survey}. Fortunately, visual sensing information captured by base station (BS) cameras can be leveraged to aid beam prediction, providing rich environmental awareness such as vehicle locations, mobility patterns, and blockage conditions that are highly correlated with the optimal communication beams. 
This sensing-aided approach eliminates the need for exhaustive beam search, enabling proactive beam management within the stringent latency budget imposed by vehicular mobility.

\begin{figure}[!t]
\centering
\includegraphics[width=0.82\columnwidth]{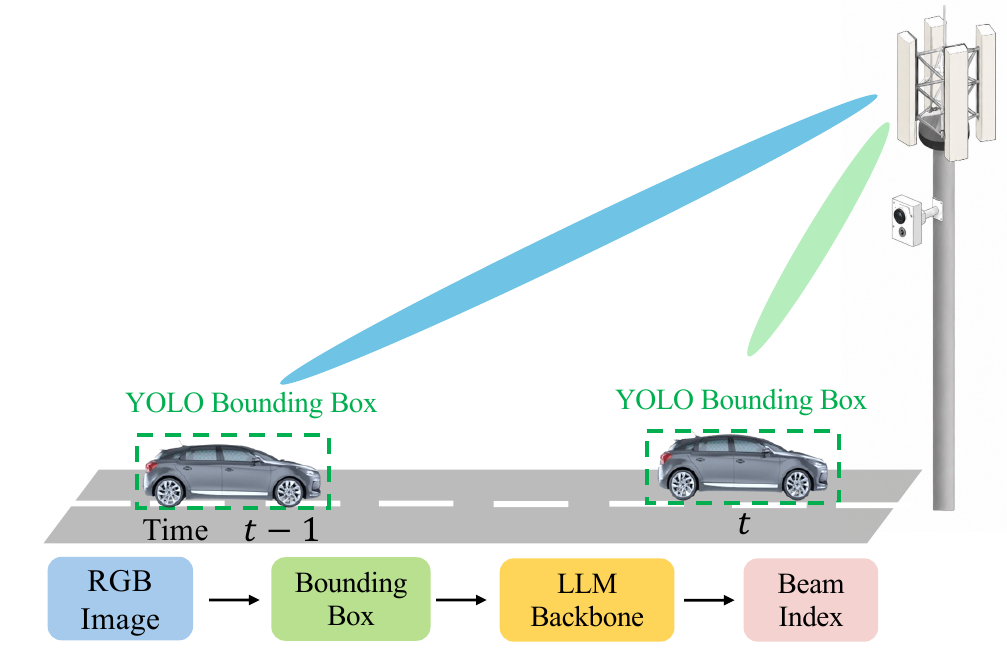}
\caption{System overview of vision-aided  V2I beam prediction. A BS equipped with a phased array and an RGB camera serves a moving vehicle, and the captured bounding-box trajectory is processed by LLM backbone to predict the beam index.}
\label{fig:system}
\end{figure}

Early approaches employ lightweight recurrent models to predict future beams from camera-captured vehicle trajectories, such as recurrent neural network (RNN)~\cite{Jiang2022CV}, gated recurrent units (GRU) \cite{ma2025knowledge}, and long short-term memory networks (LSTM) \cite{zarei2025lstm}. While these models are computationally efficient, they learn representations entirely from task-specific training data and therefore degrade significantly in limited-data settings. More recently, BeamLLM~\cite{beamllm} addresses this limitation by reprogramming a pretrained GPT-2 model~\cite{radford2019gpt2} for beam prediction, achieving substantially improved performance. However, the resulting gains come at the cost of significantly increased model size and high inference latency, limiting its suitability for real-time roadside deployment. Therefore, there exists a tradeoff between data efficiency and deployment efficiency in vehicle-to-infrastructure (V2I) beam prediction, which remains largely unexplored. Moreover, it remains unclear which components of a pretrained LLM are actually responsible for the performance gains since most existing approaches retain the entire LLM architecture without examining structural redundancy for efficient deployment.

Motivated by these gaps, we propose \textbf{LLMBP-Lite}, a lightweight LLM-reprogrammed beam predictor that compresses the model while preserving pretrained representations for data-efficient and deployment-friendly V2I beam prediction. The main contributions are as follows:
\begin{itemize}
\item A three-stage compression framework is proposed based on a systematic redundancy analysis of LLM-reprogrammed beam predictors. The framework combines layer pruning, vocabulary pruning, and soft-prompt replacement to substantially 
reduce model size and inference latency. Optional knowledge distillation (KD) is further incorporated to preserve pretrained representations after compression.
\item A systematic analysis is conducted to examine whether the data efficiency of LLM-reprogrammed beam prediction is preserved under aggressive structural compression. Results show that LLMBP-Lite consistently outperforms recurrent baselines under limited training data, revealing that pretrained representation remains effective after compression.
\item Comprehensive experiments on the DeepSense\,6G dataset~\cite{deepsense2022} are conducted to evaluate LLMBP-Lite against both LLM-based and recurrent baselines across accuracy, model size, and inference latency. The results demonstrate a $65\times$ inference speedup over the uncompressed BeamLLM baseline with an optimal 
accuracy--latency tradeoff for practical V2I deployment.

\end{itemize}

\begin{figure*}[!t]
  \centering
  \includegraphics[width=0.8\textwidth]{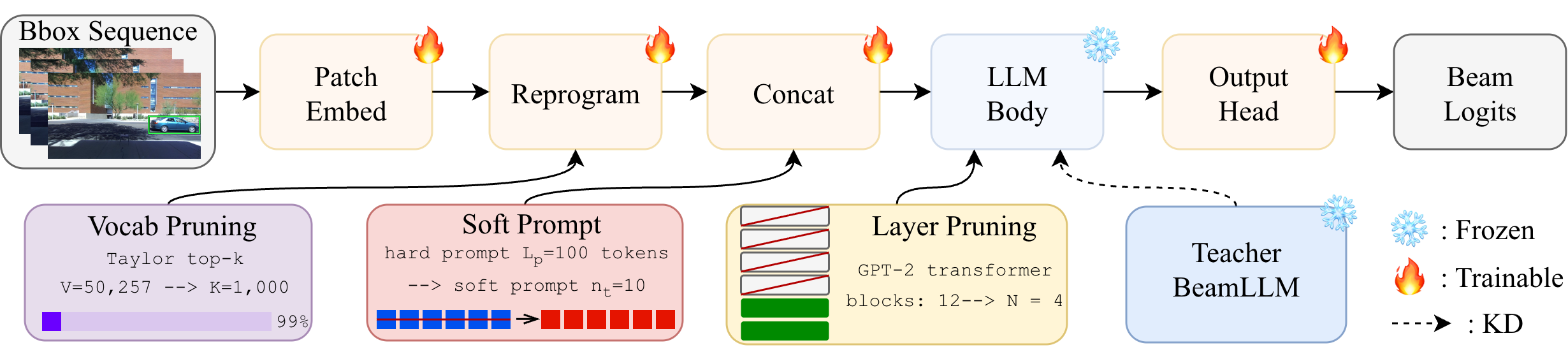}
  \caption{Proposed LLMBP-Lite. The original BeamLLM is compressed along three complementary dimensions: Transformer depth (\textit{Layer Pruning}), source-prototype vocabulary size (\textit{Vocab Pruning}), and prompt length (\textit{Soft Prompt}). The dashed arrow denotes optional knowledge distillation from the original BeamLLM teacher. Snowflake and fire icons indicate frozen and trainable components, respectively.}
  \label{fig:arch}
\end{figure*}

\section{Proposed LLMBP-Lite}
\label{sec:method}

As illustrated in Fig.~\ref{fig:arch}, three compression stages are applied in a coarse-to-fine pipeline. Specifically, layer pruning first reduces the backbone depth, vocabulary pruning then selects the most informative source prototypes relative to the compressed backbone, and soft-prompt replacement finally shortens the conditioning sequence. Moreover, KD from the original teacher model is optionally incorporated to recover pretrained representations lost during compression, thereby preserving the data efficiency that motivates the use of LLM reprogramming for beam prediction in vehicular systems. The following subsections detail each 
stage.

\subsection{Depth Compression via Layer Pruning}
\label{sec:c3}
The Transformer layer stack is the primary contributor to LLM's inference latency, yet not all layers contribute equally to beam-prediction performance. 
Motivated by this observation, we construct a compressed backbone by retaining only the first $N$ ($< 12$) Transformer blocks of the pretrained GPT-2~\cite{radford2019gpt2} backbone, discarding the remaining $12 - N$ upper layers. The retained layers are directly initialized from the original pretrained GPT-2~\cite{radford2019gpt2} model, preserving the hierarchical representations learned during pretraining without any additional retraining cost. From a complexity perspective, the self-attention computation across the Transformer stack scales as $\mathcal{O}\!\left(N L^2 d\right)$ where $L$ denotes the total input sequence length and $d$ is the hidden dimension of the backbone. Thereby, reducing the number of Transformer blocks from 12 to $N$ ($< 12$) yields a nearly linear reduction in both inference latency and memory consumption, making layer pruning the most impactful compression stage in the proposed LLMBP-Lite.

\subsection{Prompt Compression via Soft Prompting}
\label{sec:c4}
Recent LLM-based beam prediction work in \cite{beamllm} conditions the frozen GPT-2~\cite{radford2019gpt2} backbone with a handcrafted hard prompt of approximately $L_p \approx 100$ tokens. While effective in steering the frozen backbone toward beam prediction, this prompt directly increases the sequence length processed by every Transformer layer. This amplifies the quadratic $\mathcal{O}(L^2)$ 
attention cost, consequently raising both inference 
latency and memory overhead. 
Motivated by this observation, we replace the hard prompt with a compact learnable soft prompt $\mathbf{S} \in \mathbb{R}^{n_t \times d}$, where $n_t \ll L_p$. Rather than random initialization, $\mathbf{S}$ is initialized from the token embeddings of the original hard prompt, allowing the soft prompt to retain the semantic information 
encoded in the original hard prompt and accelerate fine-tuning (FT) convergence. During FT, $\mathbf{S}$ is the only prompt-side parameter updated, while the GPT-2~\cite{radford2019gpt2} backbone remains frozen. This replacement reduces the effective sequence length from
$L = L_p + Q$ to $L' = n_t + Q$, where $Q$ denotes the number of reprogrammed patch tokens. This yields a significant reduction in attention cost owing to its quadratic dependence on sequence length. Beyond the computational benefit, replacing discrete token indices with continuous embeddings can also eliminate prompt tokenization overhead during deployment, further reducing inference latency on resource-constrained BSs.

\subsection{Vocabulary Compression via Taylor Pruning}
\label{sec:c2}
The source-prototype mapping matrix $\mathbf{W}^{\text{map}} \in 
\mathbb{R}^{d \times V}$ is the largest non-Transformer component in LLM-based beam prediction model, where $V$ denotes the full GPT-2~\cite{radford2019gpt2} vocabulary size.  During reprogramming, the patch embeddings are aligned 
with the columns of $\mathbf{W}^{\text{map}}$ through 
a cross-attention mechanism to acquire language-aligned representations.  However, not all $V$ source prototypes contribute equally to this process: a large fraction of vocabulary entries receive negligible attention weights and can therefore be removed without affecting prediction performance.
To identify the most informative entries, we introduce
the importance score of vocabulary entry $v \in [1, \cdots, V]$, which is defined as ~\cite{molchanov2017pruning}
\begin{equation*}
  I(v) \;=\; \mathbb{E}
  \Big[\,\left\|\;
\nabla_{\mathbf{W}^{\text{map}}_{[:,v]}}\,\mathcal{L}_{\text{CE}}
\;\odot\;\mathbf{W}^{\text{map}}_{[:,v]}\;\right\|_1\,\Big],
\end{equation*}
where $\nabla_{\mathbf{W}_{[:,v]}^{\text{map}}}\,\mathcal{L}_{\text{CE}}$ denotes 
the gradient of the cross-entropy loss with respect to the $v$-th column of $\mathbf{W}^{\text{map}}$, and $\odot$ denotes element-wise multiplication. This score indicator captures the joint effect of parameter magnitude and gradient sensitivity, and effectively quantifies the contribution of each source prototype to the reprogramming process.

The importance scores of all vocabulary entries are collected 
into a vector $\mathbf{a} = [I(v)]_{v=1}^{V}$, and the $L$ entries with the highest importance scores among $\mathbf{a}$ (i.e., TopL$(\mathbf{a},\, L)$) are 
selected to form a retained index set $\mathcal{I}$:
\begin{equation*}
    \mathcal{I} = \operatorname{TopL}(\mathbf{a},\, L), \quad
    \mathbf{W}^{\text{map}} \leftarrow \mathbf{W}_{[:,\,\mathcal{I}]}^{\text{map}},
\end{equation*}
where $|\mathcal{I}| = L$ and $\mathbf{W}_{[:,\,\mathcal{I}]}^{\text{map}}$ denotes the submatrix formed by retaining only the columns indexed by $\mathcal{I}$. Following pruning, the model is fine-tuned on $\mathcal{D}_{\text{tr}}$ 
to allow the remaining components to adapt to the reduced 
source-prototype vocabulary before subsequent compression stages.

\subsection{Knowledge Distillation for Representation Preservation}
\label{sec:kd}
While the three compression stages significantly reduce the computational cost, aggressive pruning inevitably discards some pretrained representations that contribute to its data efficiency, especially in limited-data regimes. To compensate for this loss, we incorporate KD from the original LLM-based model as the teacher model $f_{\rm T}$ into the 
compressed LLMBP-Lite as the student model $f_{\rm S}$.
Rather than treating KD purely as a compression technique, we employ it specifically as a representation-preservation 
mechanism. The student model $f_{\rm S}$ is trained to mimic the soft output distribution of the frozen teacher $f_{\rm T}$, encouraging the compressed model to recover the predictive behavior shaped by GPT-2~\cite{radford2019gpt2} pretraining. 

Let $\mathbf{p}_{\rm S} = f_{\rm S}(\mathbf{x}) \in \mathbb{R}^B$ and 
$\mathbf{p}_{\rm T} = f_{\rm T}(\mathbf{x}) \in \mathbb{R}^B$ denotes the predicted beam logits of the student model and teacher model, respectively, where $B$ is the codebook size. The student model is optimized using a composite loss:
\begin{align*}
\mathcal{L}_{\text{KD}}= -\log\frac{\exp(\mathbf{p}_{\rm S}^{(y)})}
{\displaystyle\sum_{b=1}^{B}\exp(\mathbf{p}_{\rm S}^{(b)})} +\alpha T^2\mathrm{KL}\!\left[\,\sigma\!\left(\frac{\mathbf{p}_{\rm T}}{T}\right)
\,\|\,\sigma\!\left(\frac{\mathbf{p}_{\rm S}}{T}\right)\right],
  \label{eq:kd}
\end{align*}
where $y \in \{1, \ldots, B\}$ denotes the ground-truth optimal beam index; 
$\mathrm{KL}[\cdot\|\cdot]$ is the Kullback--Leibler (KL) divergence~\cite{ma2025knowledge}; $\sigma(\cdot)$ denotes the softmax function, $T$ is the distillation temperature, and $\alpha$ balances the two loss terms. The first term of $\mathcal{L}_{\text{KD}}$ expression is the cross-entropy loss that supervises the student with ground-truth beam labels, while the second term is the KL divergence that penalizes deviations of the student's soft output distribution from that of the teacher, encouraging the compressed model to mimic the predictive behavior shaped by GPT-2~\cite{radford2019gpt2} pretraining.

As illustrated in Fig.~\ref{fig:arch}, KD is applied as the final stage of the proposed LLMBP-Lite and is optionally enabled depending on the deployment scenario. When activated, it transfers the data-efficiency characteristics of the original LLM-based model to the compressed 
student, ensuring that LLMBP-Lite retains its advantage over conventional baselines even under limited training data, which will be validated in the following section.

\begin{table*}[!t]
\centering
\caption{Performance comparison on Scenario~8 of DeepSense\,6G~\cite{deepsense2022}  
with $H=8$, $P=5$, and $n=300$. Top-$k$ accuracies are averaged over the $P=5$ prediction steps. Speedup is computed relative to the uncompressed BeamLLM baseline.
}
\label{tab:main}
\renewcommand{\arraystretch}{1.15}
\setlength{\tabcolsep}{4pt}
\renewcommand{\arraystretch}{0.9}
\begin{tabular}{l ccc c c c c  c}
\toprule
 Method & Top-1 & Top-3 & Top-5 & \# of parameters & Size (MB) & FLOPs (G) & GPU latency (ms)  & Speed-up \\
\midrule
RNN                       & 0.482 & 0.858 & 0.950 & 69\,k & 0.3 & $<\!0.01$ & 0.013 & 1439$\times$ \\
GRU                       & 0.512 & 0.882 & 0.969 & 170\,k & 0.6 & $<\!0.01$ & 0.008  & 2286$\times$ \\
LSTM                      & 0.488 & 0.867 & 0.966 & 220\,k & 0.8 & $<\!0.01$ & 0.065  & 279$\times$ \\
\midrule
BeamLLM \cite{beamllm} (uncompressed, no FT) & 0.548 & 0.944 & 0.996 & 178.3\,M & 692.2 & 136 & 18.00  & 1.0$\times$ \\
+ Layer pruning ($N{=}4$) & 0.575 & 0.952 & 0.999 & 121.6\,M & 467.9 & 72.8 & 7.197  & 2.5$\times$ \\
+ Soft prompt ($t{=}10$)  & 0.576 & 0.951 & 0.999 & 178.3\,M & 692.2 & 44.9 & 0.895  & 20$\times$ \\
+ Layer pruning + KD ($N{=}4$) & 0.576 & 0.951 & 0.998 & 121.6\,M & 467.9 & 72.8 & 7.197 & 2.5$\times$ \\
BeamLLM + FT & 0.574 & 0.951 & 0.999 & 178.3\,M & 692.2 & 136 & 18.00 & 1.0$\times$ \\
\midrule
LLMBP-Lite (ours) & 0.573 & 0.951 & 0.999 & 71.2\,M & 278.4 & 3.66 & 0.276 & 65$\times$ \\
\bottomrule
\end{tabular}
\end{table*}

\section{Experiments}
\label{sec:experiments}

We evaluate the system performance based on Scenario~8 of the DeepSense 6G~\cite{deepsense2022} dataset, a V2I outdoor scenario with a 64-beam codebook. Specifically,  YOLO~\cite{wang2023yolov7} is adopted to extract vehicle bounding boxes from RGB images, and the received power profiles are collected by a $60$ GHz phased antenna array. To evaluate performance under limited training data, we vary the number of training samples $n$ while fixing $300$ samples for validation, and $600$ samples for testing. Each model takes the most recent $H = 8$ bounding boxes as input and predicts the next $P=5$ beam indices. 

We compare LLMBP-Lite against two groups of baselines: the BeamLLM proposed in~\cite{beamllm} as the uncompressed upper bound, and three classical recurrent baselines, namely RNN, GRU, and LSTM. LLM-family models are initialized from a pretrained BeamLLM~\cite{beamllm} checkpoint trained for $200$ epochs and subsequently fine-tuned for an additional $100$ epochs. The default LLMBP-Lite configuration uses $N=4$ Transformer layers, $K=1000$ source-prototype entries, and a soft prompt of length $n_t=10$, without KD.
All experiments are conducted on a single NVIDIA A40 GPU with a batch size of $16$. All models are optimized using the Adam optimizer with a learning rate of $10^{-3}$. 
Performance is evaluated using Top-$k$ beam prediction accuracy ($k \in \{1, 3, 5\}$), averaged over the $P$ prediction horizons. Top-1 accuracy serves as the primary metric, while Top-3 and Top-5 are reported for completeness.

\begin{figure}[!t]
  \centering
  \includegraphics[width=0.65\columnwidth]{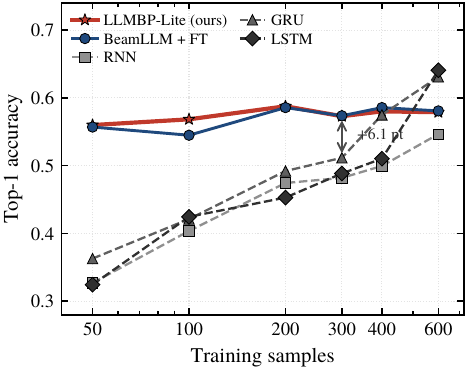}
 \caption{Top-1 beam prediction accuracy versus training-set 
size $n$ for LLMBP-Lite, uncompressed BeamLLM, and 
recurrent baselines.} 
  \label{fig:sample}
\end{figure}

Fig.~\ref{fig:sample} demonstrates data-efficiency advantage of LLM-reprogrammed beam prediction after aggressive compression by varying the training-set size $n \in \{50, 100:100:400, 600\}$. 
Specifically, the recurrent baselines benefit strongly from additional training data, with Top-1 accuracy improving from approximately $0.32$-$0.36$ at $n =50$ to $0.55$-$0.64$ at $n=600$. In contrast, LLMBP-Lite remains stable within $0.560$-$0.588$ across the entire range $n$, closely approaching the uncompressed BeamLLM ($0.545$-$0.586$), indicating that compression preserves the sample-efficiency behavior of the original model. A crossover is observed near $n\approx 400$. Below this threshold, LLMBP-Lite consistently outperforms all recurrent baselines. The Top-1 accuracy increases from $0.5$ percentage points (pts) at $n=400$ to $6.1$ pts at $n=300$ and $19.7$ pts at $n=50$, demonstrating a significant advantage under a limited data region. Above this threshold, task-specific recurrent models can more fully exploit the abundant training data. Based on these findings, all subsequent experiments are conducted 
under the limited-data setting of $n = 300$.

\label{sec:main-results}
Table~\ref{tab:main} summarizes prediction accuracy, model 
complexity, and inference latency at the limited-data operating point $n =300$. LLMBP-Lite achieves a Top-1 accuracy of $0.573$, outperforming the strongest GRU baseline ($0.512$) by $6.1$ pts and the uncompressed BeamLLM without FT ($0.548$) by $2.5$ pts. Furthermore, LLMBP-Lite reduces model size from $692.2$\,MB to $278.4$\,MB, parameter count from $178.3$\,M to $71.2$\,M, FLOPs from $136$\,G to $3.66$\,G, and GPU latency from $18.0$\,ms to $0.276$\,ms, corresponding to a $65\times$ inference speedup. It achieves the best accuracy-latency trade-off among all evaluated configurations. 
The BeamLLM+FT scheme achieves a Top-1 accuracy of $0.574$, 
essentially identical to LLMBP-Lite with accuracy of $0.573$. This demonstrates that the performance gap between the two is negligible despite the $65\times$ reduction in inference latency and significant reductions in model size and FLOPs. Finally, Top-3 and Top-5 accuracies are comparable and remain consistently high across all BeamLLM-based models, indicating that the proposed compression scheme preserves the performance of the original BeamLLM.
\begin{table}[t]
\centering
\caption{Component-wise ablation of LLMBP-Lite.}
\label{tab:ablation}
\renewcommand{\arraystretch}{1.2}
\setlength{\tabcolsep}{4pt}
\renewcommand{\arraystretch}{0.9}
\begin{tabular}{l c cc c r r}
\toprule
Configuration & Vocab. & Layers & Prompt & Top-1 & \makecell{Size \\ (MB)}& \makecell{Latency \\(ms)}\\
\midrule
BeamLLM \cite{beamllm} &   &     &   & 0.548 & 692.2 & 18.00 \\
+ Vocab & \checkmark &   &    & 0.572 & 502.7 & 41.80 \\
+ Layers &   & \checkmark &     & 0.575 & 467.9 & 7.197 \\
+ Prompt &   &   & \checkmark  & 0.576 & 692.2 & 0.895 \\
+ Layers + KD  &   & \checkmark &   & 0.576 & 467.9 & 7.197 \\
+ Vocab + Layers & \checkmark & \checkmark &   & 0.566 & 278.4 & 20.73 \\
+ Layers + Prompt  &   & \checkmark & \checkmark & 0.573 & 467.9 & 0.398 \\
LLMBP-Lite & \checkmark & \checkmark & \checkmark   & 0.573 & 278.4 & 0.276 \\
+ KD & \checkmark & \checkmark & \checkmark & 0.576 & 278.4 & 0.276 \\
\bottomrule
\end{tabular}
\end{table}
To isolate the contribution of FT from that of compression, we compare three configurations from Table~\ref{tab:main} at $n = 300$: the released BeamLLM checkpoint (Top-1 $=0.548$)~\cite{beamllm}, the fine-tuned BeamLLM without compression (Top-1 $=0.574$), and 
LLMBP-Lite (Top-1 $=0.573$). The results show that FT alone provides the majority of the accuracy gain, increasing Top-1 from $0.548$ to $0.574$ ($+2.6$\,pt). The subsequent compression reduces Top-1 by only $0.001$\,pt 
from $0.574$ to $0.573$ despite reducing model size from $692.2$\,MB to $278.4$\,MB and GPU latency from $18.0$\,ms to $0.276$\,ms. 

Table~\ref{tab:ablation} further examines the interaction among the compression stages, with all configurations evaluated at the default settings ($K = 1000$, $N = 4$, $n_t =10$). Specifically, all single-stage configurations achieve Top-1 accuracies of $0.566$-$0.576$, closely matching the fine-tuned BeamLLM ($0.574$) while reducing either model size or latency. This demonstrates that once the model is fine-tuned under the proposed scheme, the specific combination of compression stages has little effect on Top-1 accuracy while strongly affecting computational cost. Notably, prompt compression alone reduces GPU latency from $18.0$\,ms to $0.895$\,ms ($20\times$ speedup) with negligible accuracy change, making it the single most impactful stage for latency reduction. Layer pruning contributes the largest reduction in model size ($692.2$ to $467.9$\,MB). Combining all three stages in LLMBP-Lite achieves the smallest model size ($278.4$\,MB) and lowest GPU latency ($0.276$\,ms) with a Top-1 accuracy of $0.573$, which is comparable to all single-stage variants. Furthermore, applying KD on LLMBP-Lite increases Top-1 accuracy from $0.573$ to $0.576$.
This shows that KD acts as an effective regularizer in the limited-data regime, where the trainable components introduced by compression are optimized from relatively few samples.

\begin{figure*}[!t]
  \centering
  \includegraphics[width=0.72\textwidth]{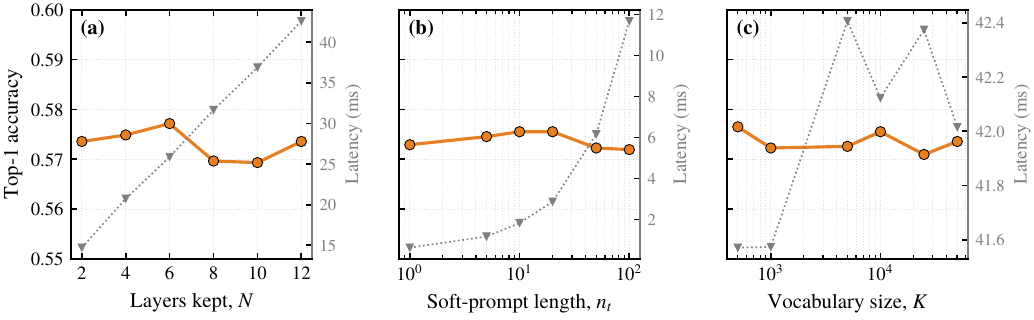}
\caption{Impact of individual compression stages at 
$n =300$. Top-1 accuracy (left axis) and per-sample 
GPU latency (right axis) are shown for each compression 
dimension: (a) layer pruning over retained depth $N\in\{2,4,6,8,10,12\}$; 
(b) soft prompting over prompt length  $n_t \in\{1,5,10,20,50,100\}$; and (c) vocabulary pruning over retained vocabulary size $K \in \{500,\ldots,50 257\}$.}
\label{fig:sweep}
\end{figure*}

Fig.~\ref{fig:sweep} shows the sensitivity of the proposed LLMBP-Lite to each compression dimension independently at $n = 300$. A consistent trend is observed across all three dimensions: Top-1 accuracy remains stable within a narrow band of approximately $0.57$ despite a substantial variation range in model structure and inference cost. This demonstrates that predictive performance is largely insensitive to the degree of compression. Furthermore, those three compression dimensions exhibit complementary effects on inference efficiency, as shown in 
Fig.~\ref{fig:sweep}. Specifically, layer pruning reduces latency by removing Transformer computation, with GPU latency 
decreasing from approximately $43$\,ms at $N =12$ to 
approximately $15$\,ms at $N =2$, as illustrated in Fig.~\ref{fig:sweep}(a). Soft prompting provides a large latency reduction by shortening the effective input sequence, as shown in Fig.~\ref{fig:sweep}(b). The latency remains low for small $n_t$ but increases substantially as 
the prompt length grows.  In contrast, vocabulary 
pruning has a negligible effect on latency since source 
prototypes are only accessed during the reprogramming 
stage, as shown in Fig.~\ref{fig:sweep}(c).  Its primary benefit is a reduction in model size.  These complementary properties explain why combining all three stages yields the most favorable efficiency profile in 
Table~\ref{tab:ablation}.
It is noted that Top-1 accuracy remains within a narrow band of approximately $0.57$ across the full range of each 
compression dimension, indicating that LLMBP-Lite 
does not require careful tuning of individual compression parameters. Substantial reductions in model depth, prompt length, and vocabulary size all incur little or almost no accuracy loss. This allows flexible adaptation of the compression configuration to different deployment constraints such as stricter latency requirements or tighter memory budgets.

\begin{figure}[!t]
  \centering
\includegraphics[width=0.715\columnwidth]{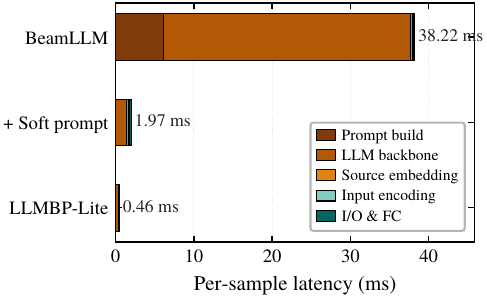}
\caption{Inference latency analysis of the forward 
pass for the uncompressed BeamLLM, the soft-prompt 
variant, and LLMBP-Lite.
}
\label{fig:latency}
\end{figure}

To understand how the $65\times$ GPU speedup is achieved, 
Fig.~\ref{fig:latency} presents a per-region latency analysis of the forward pass across three configurations: 
the uncompressed BeamLLM, the soft-prompt variant, and LLMBP-Lite.
As shown in Fig.~\ref{fig:latency}, the latency of the uncompressed BeamLLM is concentrated in two components: the GPT-2~\cite{radford2019gpt2} backbone accounts for $82.7\%$ of the total 
measured latency, while prompt construction contributes 
a further $16.0\%$. These two stages consume $98.7\%$ of the total latency budget. 
This analysis explains the effectiveness of the proposed compression scheme. Replacing the approximately $100$-token hard prompt with a compact soft prompt of length $n_t = 10$ eliminates the prompt-construction stage entirely and shortens the input sequence. This reduces latency from $38.22$\,ms to $1.97$\,ms, corresponding to a $19.4\times$ speedup. Subsequent layer pruning further reduces Transformer computation, yielding the final LLMBP-Lite latency of $0.46$\,ms. Notably, the majority of the speedup comes from eliminating prompt-related overhead rather than from reducing Transformer depth, identifying prompt processing as the primary latency bottleneck in the original BeamLLM scheme. 

\section{Conclusion}
\label{sec:conclusion}
This paper proposed LLMBP-Lite, a compressed LLM-reprogrammed beam predictor that combines layer pruning, 
vocabulary pruning, soft prompting, and knowledge distillation 
to reduce deployment cost while preserving the data-efficiency advantage of LLM reprogramming. Experiments demonstrate that LLMBP-Lite achieves   $65\times$ faster inference latency than the original BeamLLM, while consistently outperforming conventional recurrent baselines under limited training data. These results show that the computational overhead of BeamLLM can be aggressively reduced without sacrificing its low-data generalization capability, enabling practical real-time beam prediction in resource-constrained vehicular networks. 

\bibliographystyle{IEEEtran}
\bibliography{IEEEabrv,refs}  

@STRING{IEEE_M_COM        = "{IEEE} Commun. Mag."}

@STRING{jan 	   = "Jan."}

@STRING{may		   = "May"}

@STRING{JAN		   = "Jan."}

@STRING{MAY		   = "May"}

@ARTICLE{deepsense2022,
  author={Alkhateeb, Ahmed and Charan, Gouranga and Osman, Tawfik and Hredzak, Andrew and Morais, Joao and Demirhan, Umut and Srinivas, Nikhil},
  journal=IEEE_M_COM, 
  title={Deep{S}ense {6G}: A Large-Scale Real-World Multi-Modal Sensing and Communication Dataset}, 
  month={Sep.},
  year={2023},
  volume={61},
  number={9},
  pages={122-128},
  publisher={IEEE}
  }

@article{dong2025network,
  title={Network-Controlled Repeater Aided Time-Sensitive Communications in Urban Vehicular Networks},
  author={Dong, Kai and Yu, Hao and Taleb, Tarik and Sezgin, Aydin and Vorobyov, Sergiy A},
  journal={IEEE Wireless Commun. Lett.},
  month={May},
  year={2025},
  volume={14},
  number={5},
  pages={1511--1515},
  publisher={IEEE}
}

@article{ma2025knowledge,
  title={Knowledge Distillation for Sensing-Assisted Long-Term Beam Tracking in mm{W}ave Communications},
  author={Ma, Mengyuan and Nguyen, Nhan Thanh and Shlezinger, Nir and Eldar, Yonina C and Swindlehurst, A Lee and Juntti, Markku},
  journal={arXiv preprint arXiv:2509.11419},
  year={2025}
}

@article{radford2019gpt2,
  author={Radford, Alec and Wu, Jeffrey and Child, Rewon and Luan, David and Amodei, Dario and Sutskever, Ilya},
  title={Language Models are Unsupervised Multitask Learners},
  journal={OpenAI Blog},
  year={2019},
  volume={1},
  number={8},
  month={Feb.}
}

@inproceedings{zarei2025lstm,
  title={{LSTM}-based Beam Prediction for {6G} Drone Communications: Image vs. {GPS}-based Sensing},
  author={Zarei, Zahra and Tilahun, Fitsum D and Kang, Chung G},
  booktitle={Proc. 30th Asia-Pacific Conf. Commun. (APCC)},
  pages={1--5},
  year={2025},
  address={Osaka, Japan},
  month={Dec.}
}

@article{xue2024survey,
  title={A survey of beam management for {mmWave} and {THz} communications towards {6G}},
  author={Xue, Qing and Ji, Chengwang and Ma, Shaodan and Guo, Jiajia and Xu, Yongjun and Chen, Qianbin and Zhang, Wei},
  journal={IEEE Commun. Surv. Tutorials},
  volume={26},
  number={3},
  pages={1520--1559},
  year={2024},
  month={Feb.},
  publisher={IEEE}
}

@INPROCEEDINGS{Jiang2022CV,
  author={Jiang, Shuaifeng and Alkhateeb, Ahmed},
  booktitle={Proc. IEEE Globecom
  Workshops}, 
  title={Computer Vision Aided Beam Tracking in A Real-World Millimeter Wave Deployment}, 
  year={2022},
  month={Dec.},
  address={Rio de Janeiro, Brazil},
  pages={142-147}}

@inproceedings{beamllm,
  title={Beam{LLM}: Vision-empowered mm{W}ave beam prediction with large language models},
  author={Zheng, Can and He, Jiguang and Cai, Guofa and Yu, Zitong and Kang, Chung G},
  booktitle={Proc. IEEE 102nd Veh. Technol. Conf. (VTC2025-Fall)},
  pages={1--6},
  year={2025},
  month={Oct.},
  address={Chengdu, China}
}

@inproceedings{molchanov2017pruning,
    title={Pruning Convolutional Neural Networks for Resource Efficient Inference},
    author={Molchanov, Pavlo and Tyree, Stephen and Karras, Tero and Aila, Timo and Kautz, Jan},
    booktitle={Int. Conf. Learn. Represent.},
    year={2017}
}

@inproceedings{wang2023yolov7,
  title={{YOLO}v7: Trainable bag-of-freebies sets new state-of-the-art for real-time object detectors},
  author={Wang, Chien-Yao and Bochkovskiy, Alexey and Liao, Hong-Yuan Mark},
  booktitle={Proc. IEEE Conf. Comput. Vis. Pattern Recognit.},
  pages={7464--7475},
  year={2023}
}

\end{document}